\documentclass[aps,nofootinbib,twocolumn,notitlepage,floatfix,superscriptaddress,secnumarabic]{revtex4-1}

\usepackage{graphicx,color}
\usepackage{multirow}
\usepackage{times}
\usepackage{bm}
\usepackage{tabularx}
\usepackage{epstopdf}
\usepackage{enumerate}
\usepackage[normalem]{ulem}

\begin{document}

\title{
A fast centrality-meter for heavy-ion collisions at the CBM experiment
}
\author{Manjunath Omana Kuttan }
\email{manjunath@fias.uni-frankfurt.de}
\affiliation{Frankfurt Institute for Advanced Studies, 
D-60438 Frankfurt am Main, Germany}

\affiliation{Institut f\"ur Theoretische Physik,
 Johann Wolfgang Goethe Universit\"at, D-60438 Frankfurt am Main, Germany}

\author{Jan Steinheimer}
\affiliation{Frankfurt Institute for Advanced Studies, 
D-60438 Frankfurt am Main, Germany}

\author{Kai Zhou}
\email{zhou@fias.uni-frankfurt.de}
\affiliation{Frankfurt Institute for Advanced Studies, 
D-60438 Frankfurt am Main, Germany}

\author{Andreas Redelbach}
\affiliation{Frankfurt Institute for Advanced Studies, 
D-60438 Frankfurt am Main, Germany}
\affiliation{Institut f\"ur Informatik, Johann Wolfgang Goethe Universit\"at, D-60438 Frankfurt am Main, Germany}

\author{Horst Stoecker}
\affiliation{Frankfurt Institute for Advanced Studies, 
D-60438 Frankfurt am Main, Germany}
\affiliation{Institut f\"ur Theoretische Physik,
 Johann Wolfgang Goethe Universit\"at, D-60438 Frankfurt am Main, Germany}
\affiliation{GSI Helmholtzzentrum f\"ur Schwerionenforschung GmbH, D-64291
Darmstadt, Germany}
\date{\today}

\begin{abstract}
 A new method of event characterization based on Deep Learning is presented. The PointNet models can be used for fast, online event-by-event impact parameter determination at the CBM experiment. For this study, UrQMD and the CBM detector simulation are used to generate Au+Au collision events at 10 $A$GeV which are then used to train and evaluate PointNet based architectures. The models can be trained on features like the hit position of particles in the CBM detector planes, tracks reconstructed from the hits or combinations thereof. The Deep Learning models reconstruct impact parameters from 2-14~fm with a mean error varying from -0.33 to 0.22~fm.  For impact parameters in the range of 5-14~fm, a model which uses the combination of hit and track information of particles has a relative precision of 4-9 $\%$ and a mean error of -0.33 to 0.13~fm. In the same range of impact parameters, a model with only track information has a relative precision of 4-10 $\%$ and a mean error of -0.18 to 0.22~fm. This new method of event-classification is shown to be more accurate and less model dependent than conventional methods and can utilize the performance boost of modern GPU processor units.  
\end{abstract}
\maketitle

The Compressed Baryonic Matter (CBM) detector is presently being constructed for the Facility for Antiproton and Ion Research (FAIR). CBM will study the properties of strongly compressed baryonic matter using high energy nucleus-nucleus collisions with beam energies from 2 to 10 $A$GeV at the SIS-100 accelerator \cite{Friese:2006dj, Senger:2006wd, Staszel:2010zza}.
An important feature of the CBM experiment is the very high event and trigger rate which allows the detection of rare particles as well as the study of observables which require large event samples, such as higher orders of fluctuations and correlation functions. 
The full exploitation of these properties of the CBM detector requires new analysis techniques allowing for the ultra fast analysis of the continuous stream of events created at the detector.
In this work we will introduce a new analysis method based on Deep Learning (DL). In particular, we will employ this new type of model for impact parameter estimation at the CBM experiment.

Recently, there have been many applications of Machine learning (ML) techniques in high energy physics~\cite{Pang:2016vdc, Zhou:2018ill, Steinheimer:2019iso, Du:2019civ, Thaprasop:2020mzp}, and on the experimental side ML and DL methods are mainly used in tasks such as particle identification, tracking, event reconstruction and extraction of different physical observables \cite{Bourilkov:2019yoi,Radovic:2018dip,Guest:2018yhq,Larkoski:2017jix,deOliveira:2015xxd,Baldi:2016fql,Komiske:2016rsd,Almeida:2015jua,Kasieczka:2017nvn,Kasieczka:2019dbj,Qu:2019gqs,Moreno:2019bmu,Kasieczka:2020nyd,Sirunyan:2020lcu,Esmail:2019ypk,Haake:2017dpr,Samuel:2018xci,Samuel:2019crc}. These models have been shown to be superior to traditional algorithms in terms of their accuracy and processing speed. This makes machine learning techniques an ideal candidate for processing highly complex experimental data.

The impact parameter is essential in understanding the event geometry and analysis of collected data. A method which can rapidly determine the centrality of an event, even before any information on the particles created is known, would be very important for a first step event selection. In addition, an accurate determination of the initial volume of the system is very important for the analysis of fluctuations \cite{Jeon:2003gk,Skokov:2012ds} and correlations and thus for the search of observables sensitive to a possible phase transition or critical point. Although most theoretical calculations require the impact parameter as an input, it is not directly measurable in experiments. Usually, final state observables such as the mid-rapidity charged particle multiplicity and the number and energy of spectator fragments are used to determine the centrality of a collision from which then the impact parameter is estimated. For the CBM experiment, this was done using a Monte Carlo Glauber (MC-Glauber) model. These estimators are then used to group events into various centrality classes based on the centrality percentile \cite{Klochkov:2017oxr}. Note that such a method cannot determine the impact parameter of an individual event but only provides the likely distribution of impact parameters within a given centrality class.
 
 Machine Learning (ML) approaches have been previously proposed as a method for the impact parameter determination in heavy ion collisions. Feed forward networks, used with event generator output, have shown to perform better than conventional methods in \cite{Bass:1993vx,David:1994qc,Bass:1996ez}. Other studies \cite{Haddad:1996xw,DeSanctis:2009zzb,1813362} used neural networks or conventional machine learning methods to determine the impact parameter in real experimental data. However, these studies were using shallow neural networks or traditional machine learning models trained directly on output from event generators like the Quantum Molecular Dynamics (QMD), Isospin Quantum Molecular Dynamics (IQMD), Ultra relativistic Quantum Molecular Dynamics (UrQMD) or Classical Molecular Dynamics (CMD) model \cite{Aichelin:1986wa,Hartnack:1997ez,Durand:1992lzf,Charity:1988zz,Belkacem:1995zza,Dorso:1999zw}. Experimental constraints were only taken into account by simple filters based on detector acceptance or event selection criteria. Such simplifications do not take into account the uncertainties in the data introduced by detector efficiency or resolution and do not reflect the real output of a detector setup. The observables used in the previous studies are available only after several stages of processing such as track reconstruction, particle identification and efficiency corrections. Although models based on this input are easier to interpret, their main shortcoming is that any bias or constraints in the processing algorithms would also add to the uncertainty of the predictions. Furthermore, to make any judgement on the computational efficiency of such models one has to employ them in a more realistic setup that closely mimics the real data processing in the corresponding experiment.
 
Finally, an important motivation of using a neural network with direct detector output information is the flexibility of the networks output. If the analysis of the impact parameter can be done within such a model, the desired observable could be simply exchanged by other possible observable of interest and event characteristics like collective flow or the appearance of exotic particles. Thus, as we will show the DL-analysis is not only quite appropriate for the determination of the impact parameter, but serves as an example on how a generalized analysis can be performed using DL-methods.

\section{The CBM detector}\label{cbmdet}
CBM is a fixed target experiment that can be configured for electron-hadron measurements as well as muon-hadron measurements. A micro strip detector based Silicon Tracking System (STS) \cite{Heuser:2015zpa,Deveaux:2014cda} reconstructing  momenta and tracks of charged particles is one of the key components of the CBM experiment. The STS comprises of 8 equidistant planar detector stations placed from 30-100 cm downstream the target. The STS provides a single hit resolution of ~25 $\mu m$ and a momentum resolution of ~ 1$\%$. The CMOS pixel based Micro Vertex Detector (MVD) \cite{Deveaux:2014cda} is designed to reconstruct open charm decays with a secondary vertex resolution of ~50~$\mu m$. MVD comprises of 4 silicon pixel layers located 5-20 cm downstream the target. The MVD together with STS are placed in the gap of a dipole magnet with magnetic field of ~1~Tm. A Ring Imaging CHerenkov  \cite{Adamczewski-Musch:2017pix} detector is used to identify electrons from decay of low mass vector meson decay while high energy electrons and positrons are identified using the Transition Radiation Detector. Resistive Plate Chambers based Time Of Flight (TOF) measurements are used to identify hadrons \cite{herrmann2014technical}. 
Aforementioned detector systems are the basis of the electron-hadron configuration which is considered in this analysis. The collisions will produce up to 1000 charged particles at the maximum interaction rate of 10 MHz, producing ~1 Tbytes/s of raw data. The data are then processed using a First Level Event Selector (FLES) \cite{de2011first}, which performs online event building, reconstruction, tracking and event selection.
 It is interesting to note that a CBM full-system test-setup named mCBM has been constructed at the SIS18 facility of GSI/FAIR. As this setup offers additional high-rate detector tests in nucleus-nucleus collisions under realistic experimental conditions, it can be used to test the present analysis also at lower energies than at the full CBM detector.

\section{Simulation and datasets}\label{simdat}
The microscopic relativistic N-body hadron transport model UrQMD 3.4 \cite{Bass:1998ca, Bleicher:1999xi} is selected for use as event generator for the present study. UrQMD provides both a reasonable, physically well motivated scenario for the primary nucleus-nucleus collision as well as a fast, robust N-body event-by-event output in the CBM energy range. These generated UrQMD events then serve as the input to the subsequent CbmRoot \cite{root_url} detector simulation framework, which performs event-by-event transport of all particles of each event through the detector subsystems. The standard macros in CbmRoot are used to perform particle transport, detector response and event reconstruction. The default detector geometry for electron-hadron configuration (sis100$\_$electron) was simulated using the Geant3 \cite{brun1987geant} software. Since UrQMD does not include any weak or electromagnetic decays of the produced hadrons, these are performed within the Geant3 package. The present analysis includes only those particles which produce hits in the two main silicon detectors (STS and MVD). Even though the CbmRoot can perform the full detector simulation according to the experimental specifications, it does not include a realistic simulation of different backgrounds which may lead to additional noise. The study of such effects and how DL may be able to reduce the impact of detector noise, will be studied in future works.\\

 With the current simulation setup, four different datasets, labelled as \textit{Train} and \textit{Test1}-\textit{Test3}, of Au+Au collisions at 10 AGeV are generated for this study. 
The DL models were trained using dataset \textit{Train} which contains 10${^5}$ events with impact parameters in the range of 0 to 16 fm, sampled from a uniform $b$-distribution.

Datasets \textit{Test1}, \textit{Test2} and \textit{Test3} were used to quantify the performance of the trained models. 
The first testing set \textit{Test1} contains 18 subsets, each comprising of 500 events with a different but fixed impact parameter from 0 to 16 fm. Datasets \textit{Test2} and \textit{Test3} contain 10$^{6}$ and 10$^{5}$ events respectively with impact parameters sampled from a \textit{bdb} distribution (i.e. the probability of an impact parameter $b$ is proportional to $b$, from 0 - 16 fm). Thus, \textit{Test2} and \textit{Test3} contain impact parameter distributions which are different from the training set which is important for a meaningful validation of the models. Moreover, \textit{Test3} uses a modified physics scenario which will be explained later in the paper.

The features of all the datasets are presented in table \ref{table_data}.

\begin{table}[h]
\begin{tabular}{cccc}
\hline
\hline
Dataset & $\#$ events          & \begin{tabular}[c]{@{}c@{}}Impact parameter\\ $[\mathrm{fm}]$\end{tabular} & \begin{tabular}[c]{@{}c@{}}Impact parameter \\  distribution\end{tabular} \\ \hline
\textit{Train}   & 10${^5}$ & 0-16                                                               & uniform                                                                   \\ 
\textit{Test1}   & $18\times 500$              & 0.5 - 16                                           & constant                                                                  \\ 
\textit{Test2}   & 10${^6}$ & 0-16                                                               & $bdb$                                                                \\ 
\textit{Test3}   & 10${^5}$ & 0-16                                                                & $bdb$                                                                 \\ \hline
\hline
\end{tabular}
\caption{\label{table_data} Datasets used in the study. The last column defines the impact parameter distribution of the events. The training dataset has a uniform distribution of impact parameter while a constant or $bdb$ distribution is used in the testing datasets.}
\end{table}

\section{Deep Learning models} \label{dlmod}

Deep Learning is a subset of Machine Learning which uses multiple layer neural networks that can capture deep correlations in the data \cite{lecun2015deep}. This enables the computer to find better solutions to  complex problems, which traditional ML techniques cannot find. PointNet is a deep learning architecture optimised to learn from point cloud data \cite{qi2017pointnet}. Point clouds are collection of unordered points in space where each point represents the 'N' dimensional attributes of an element that contributes to the collective structure of the cloud.
One of the important features of the PointNet model is that it can learn to be invariant to the order of input points. 

The PointNet architecture can be extremely useful in nuclear and particle physics experiments, as most of the sensor or detector data has the geometrical structure of point clouds. PointNet can be used to train deep learning models which take raw experimental data as input. Here the predictions are independent of the ordering of the particle tracks or hits. In this study, we have developed four PointNet based models that learn from different types of detector outputs such as hits and tracks of particles as features to determine the impact parameter of each collision. A point in the pointcloud is therefore defined by the attributes of a hit or a track.
More detailed information on the construction and training of the PointNet can be found in the supplemental material in \ref{appendix}.

 Impact parameter regression is a supervised learning problem where the model learns to map the inputs to the impact parameter of the event upon being trained on labelled data. During training, the model goes through several samples of data to learn the correlations in the input data and the expected output. The loss function is used as a measure of how well the model has learned during the training stage. In this study, the models were trained using 75$\%$ of events in dataset \textit{Train} with the Mean Squared Error (MSE) as the loss function. The remaining 25$\%$ of events were used for validation. Other metrics such as Mean Absolute Error (MAE) and coefficient of determination ($R^{2}$) were used to select the best model for further analyses. If $y_{true}$, $y_{pred}$ and $\left< y_{true} \right>$ are the true impact parameter, DL predictions and the mean of true values respectively, the coefficient of determination is calculated as
\begin{equation}
    R^{2}=1- \frac{\sum(y_{true} - y_{pred})^{2}}{\sum(y_{true} - \left< y_{true}\right> )^{2}+\epsilon}
\end{equation}
where the second term is the fraction of variance unexplained by the predictions and $\epsilon$ is a small positive number to prevent division by zero. The sums run over all validation events.
 
 Training the models require the tuning of several hyperparameters to achieve its best performance. We started with network structures similar to the original PointNet implementation, and then tuned different hyperparameters using a trial and error method until an optimum performance, as defined by MSE, MAE and $R^2$, was observed. The models developed in this study are briefly described below.\\
 
 \noindent
\textbf{Model-1 (M-hits)}:\\
 This model (\textit{M-hits}) uses the x, y, z position of the hits of particles in the MVD detector as input attributes. Since our inputs are just hits in the detector planes, this model can perform impact parameter determination before track finding and fitting. Since the PointNet architecture requires a fixed input size, the event with maximum number of hits ($N_{max}=1995$) in the training dataset is used as reference to fix the input dimensions (N$\times$F) to be 1995$\times$3. Any event with smaller numbers of hits has the remaining rows filled with zero. When the maximum number of hits exceeded 1995 in the testing datasets, hits were dropped randomly to fit into the input dimensions. Note that in principle the input size could also be extended to take into account the exponential tail of the $N_{charge}$ distribution, but that would also increase the computational time.\\
 
  \noindent
\textbf{Model-2 (S-hits)}:\\
This model uses the x, y, z coordinates of hits in the STS detector planes. Similar to the \textit{M-hits} model, \textit{S-hits} also does not require tracking to be performed before impact parameter can be reconstructed.
The maximum number of hits present in an event in the training data was 9820. Therefore, the input dimensions (N$\times$F) were fixed to be 9820$\times$3 with provisions analogue to \textit{M-hits} to overcome smaller or larger number of hits in testing data.\\

 \noindent 
\textbf{Model-3 (MS-tracks)}:\\
The \textit{MS-tracks} model uses the features of tracks reconstructed from, both, the hits in MVD and STS, for predicting the impact parameter. Hence, this model can be used to estimate the impact parameter only after track reconstruction. In this model, the x, y, z coordinates, dx/dz, dy/dz and charge-to-momentum ratio (q/p) of tracks of particles in the first and last plane of the tracks are the attributes of a point in the 12 dimensional point cloud. Therefore, the input dimensions are 560$\times$12 (N$\times$F) where 560 is the maximum number of tracks present in an event from the training data. Events with fewer tracks are filled with rows of zeros to maintain the same input dimensionality.\\

 \noindent
\textbf{Model-4 (HT-combi)}:\\
This model learns from the combination of both hit and track information used by the \textit{M-hits} and \textit{MS-tracks} respectively. It uses the hits from MVD together with tracks reconstructed from hits in MVD and STS to determine the impact parameter of an event. It takes the MVD hits  with dimensions 1995$\times$3 and MVD + STS tracks with dimensions  560$\times$12.\\

\section{Performance of the models}\label{perf}
The DL models were trained via backpropagation until the validation  MSE (loss) started saturating or diverged from the training loss. The MAE and coefficient of determination of the validation dataset were also considered before choosing the final weights for the model. The trained models were then tested on datasets \textit{Test1}, \textit{Test2} and \textit{Test3} to evaluate their performances.  The details of the final models are tabulated in table \ref{modeldet}. All models achieved an $R^{2}$ value of about 0.98 upon training. It can be seen that increasing the complexity ($\#$ param.) increases the training duration required for the model to converge to an optimal solution. Nevertheless, all the models finally achieve similar scores for MSE, MAE and $R^{2}$ with the \textit{MS-tracks} and \textit{HT-combi} achieving a marginally better $R^{2}$ value.

\begin{table}[b]

\begin{tabular}{p{1.48cm}p{1cm}p{1.3cm}p{0.9cm}p{0.9cm}p{0.8cm}p{0.8cm}}
\hline
\hline
Model&Epochs & $\# $ param. & MSE  & MAE  & $R^{2}$ &  Events/s \\
\hline
M-hits    & 128    &$ 3 \cdot 10^{6}$          & 0.43 & 0.51 & 0.979          & 660                      \\
S-hits    & 354    &$ 3 \cdot 10^{6}$        & 0.47 & 0.54 & 0.976          & 159                      \\
MS-tracks & 372    & $ 6 \cdot 10^{6}$        & 0.40 & 0.50 & 0.981          & 1092                      \\
HT-combi  & 484    & $ 10 \cdot 10^{6}$       & 0.39 & 0.49 & 0.981          & 435   \\

\hline
\hline           
\end{tabular}
\caption{\label{modeldet} Main features of the trained DL models. An epoch is defined as a single training pass through the entire training dataset. The number of parameters ($\# $ param.) refers to the weights, biases and kernels of the model together with non-trainable parameters which define the structure of the network. This number roughly corresponds to the complexity of the model. The MSE, MAE and $R^{2}$ are for the validation data. The last column gives an estimate for the execution speed of the model on a GPU card.} 
\end{table}

 To study the speed of the DL models, 10000 events from the dataset \textit{Test2} were tested on a Nvidia Geforce RTX 2080 Ti with a graphics processing memory of 12 GB. The \textit{MS-tracks} model was found to be the fastest with a prediction speed of about 1092 events/second while the \textit{S-hits} model was the slowest with a speed of about 159 events/second. However, the \textit{MS-tracks} can only be deployed after track reconstruction, which means that some sort of pre-processing is required which takes computational time. It must also be noted that the models were not optimised for speed. It is possible to improve the model speed by reducing the model complexity, by modifying the input dimensions to make an optimum utilisation of the available resources or by using more advanced GPUs. Nevertheless, the current speed is promising to be useful for an online analysis of data, if performed parallelly on  multiple GPUs. In addition, the advantage of a more complex model, as in our study, is that it can also be used for other analysis tasks which can then be performed at a similar speed.

 \begin{figure}
    \centering
    \includegraphics[width=0.485\textwidth,height=6.05cm]{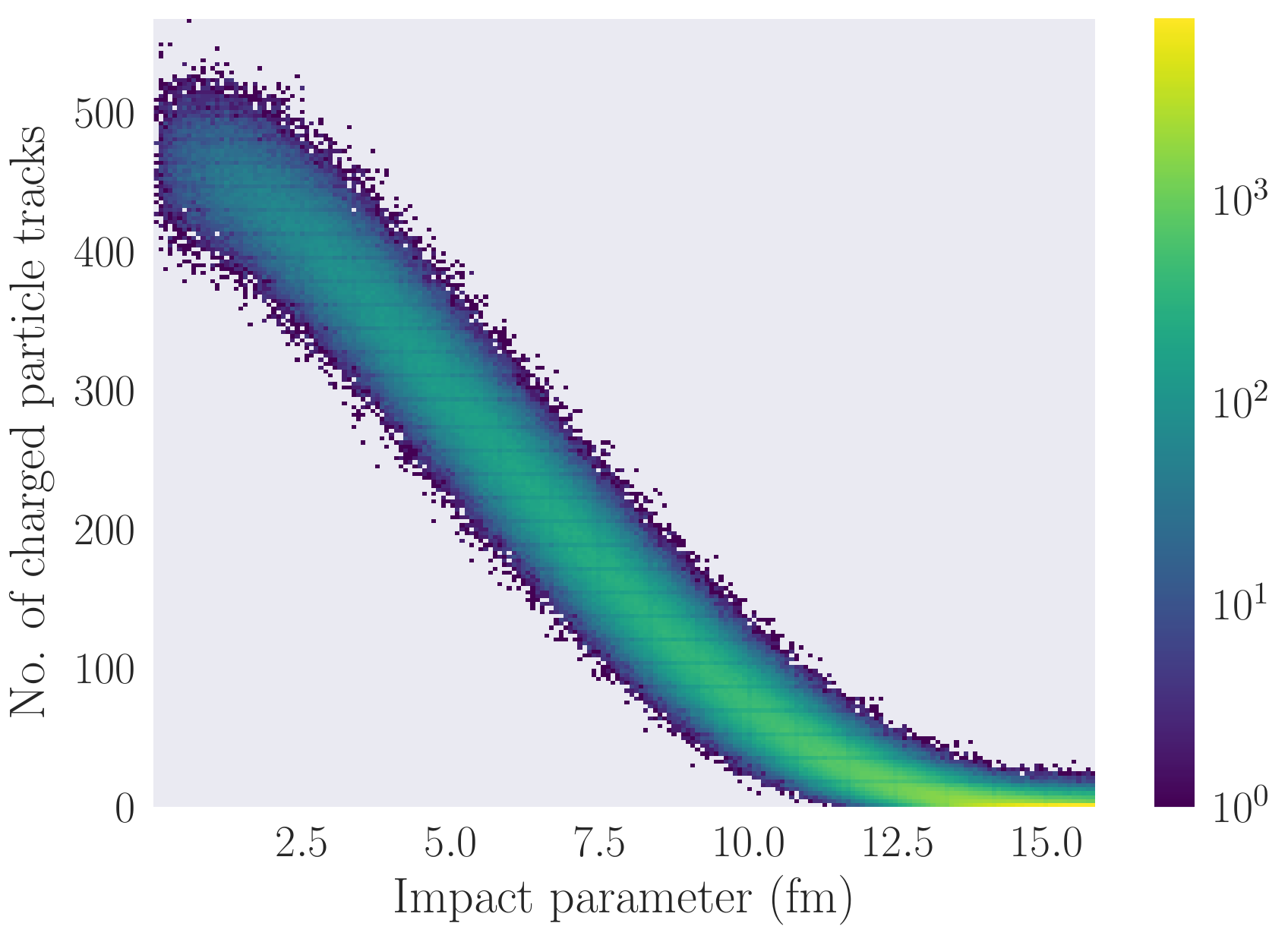}
    \caption{(Color online) Histogram of the charged particle track multiplicity as a function of impact parameter. The distribution is generated using the 10${^6}$ minimum bias events in \textit{Test2}.   }
    \label{2}
\end{figure}
While conventional methods of centrality determination, based on connecting the number of charged tracks in an event with its centrality \cite{Klochkov:2017oxr}, can be useful for a broad grouping of events, it lacks the ability to perform accurate impact parameter determination of individual events. This is evident from figure \ref{2}, in which the charged particle track multiplicity is plotted as a function of impact parameter. For a given track multiplicity, there is a wide range of possible impact parameters. This spread in track multiplicity is the largest for the most interesting central events. Similarly, for the most peripheral events, a track multiplicity could correspond to a large range of impact parameters.

\begin{figure}[t]
    \centering
    \includegraphics[width=0.485\textwidth]{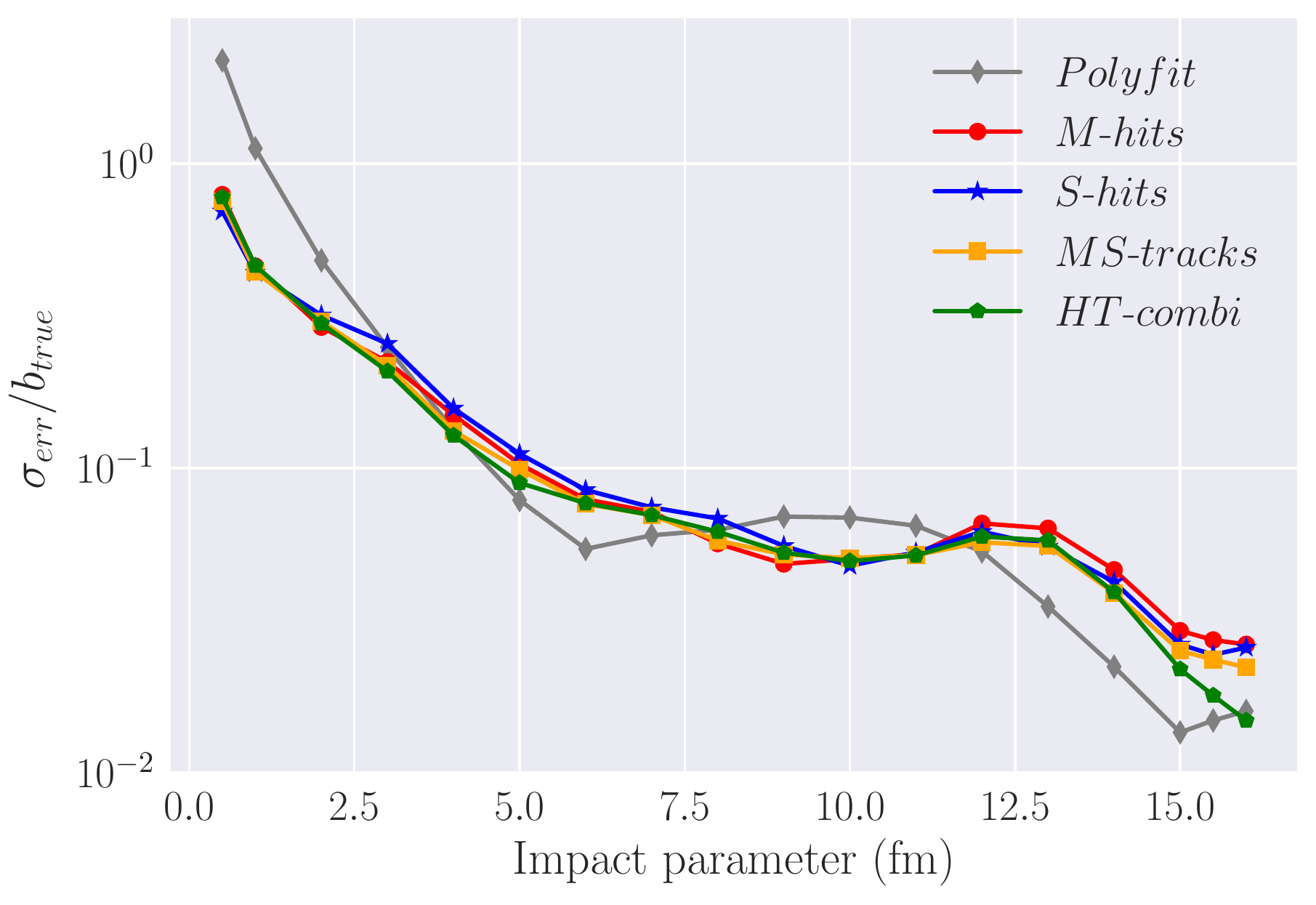}
    \caption{(Color online) Relative precision of the DL models as a function of impact parameter. The results from the $\textit{Polyfit}$ model (grey) are also plotted to benchmark the performance of DL models. The events used are from dataset $\textit{Test1}$, and predictions are for a fixed impact parameter.}
    \label{3}
\end{figure}

Accurate impact parameter determination on an event by event basis is therefore not a trivial task that can be accomplished only based on a single variable like track multiplicity. It requires modelling of other known and unknown correlations in the experimental data to the impact parameter. Moreover, an online event analysis demands minimal pre-processing of the raw experimental data. This makes PointNet based DL models an efficient candidate for event by event impact parameter determination. As a basic reference for the performance of our DL models, we will use a much simpler polynomial fit that can also perform event-by-event predictions from track multiplicity of the event. This model (\textit{Polyfit}) uses a third order polynomial fit to the track multiplicity as function of impact parameter to determine the impact parameter 
\begin{equation}
    b = a_0 + a_1\times x + a_2 \times x^{2} + a_3 \times x^{3}
\end{equation}
where $b$ and $x$ are impact parameter and the number of charged tracks, respectively. The fit gives the following parameters:\\
$a_0=14.28 $; $\ a_1=-7.01 \times 10^{-2}$; $\ a_2=2.13 \times 10^{-4}$; $ \ a_3= -2.70 \times 10^{-7}$.\\
To quantify the precision of DL models we will first look at the spread of the predictions of the DL models for a fixed input impact parameter. The relative precision in the predictions of DL models can be calculated as $\sigma_{err}/b_{true}$ where, $\sigma_{err}$ is the standard deviation of the distribution of the prediction error $(true - predicted)$ and $b_{true}$ is the true impact parameter. The relative precision in predictions is plotted as a function of impact parameter for different  DL models and the Polyfit model in figure \ref{3}. It is evident that the simple model fails for the most central collisions (b$<$ 2~fm) with the relative precision increasing up to 200 $\%$ while the DL models have a better precision in comparison. At 0.5~fm, the worst relative precision observed in DL models was about 79 $\%$ and this dropped below 50~$\%$ for events with impact parameter 1~fm or above. For events from 3 - 16~fm, the spread in  predictions of DL models and polynomial fit model are similar.

\begin{figure}
    \centering
    \includegraphics[width=0.485\textwidth]{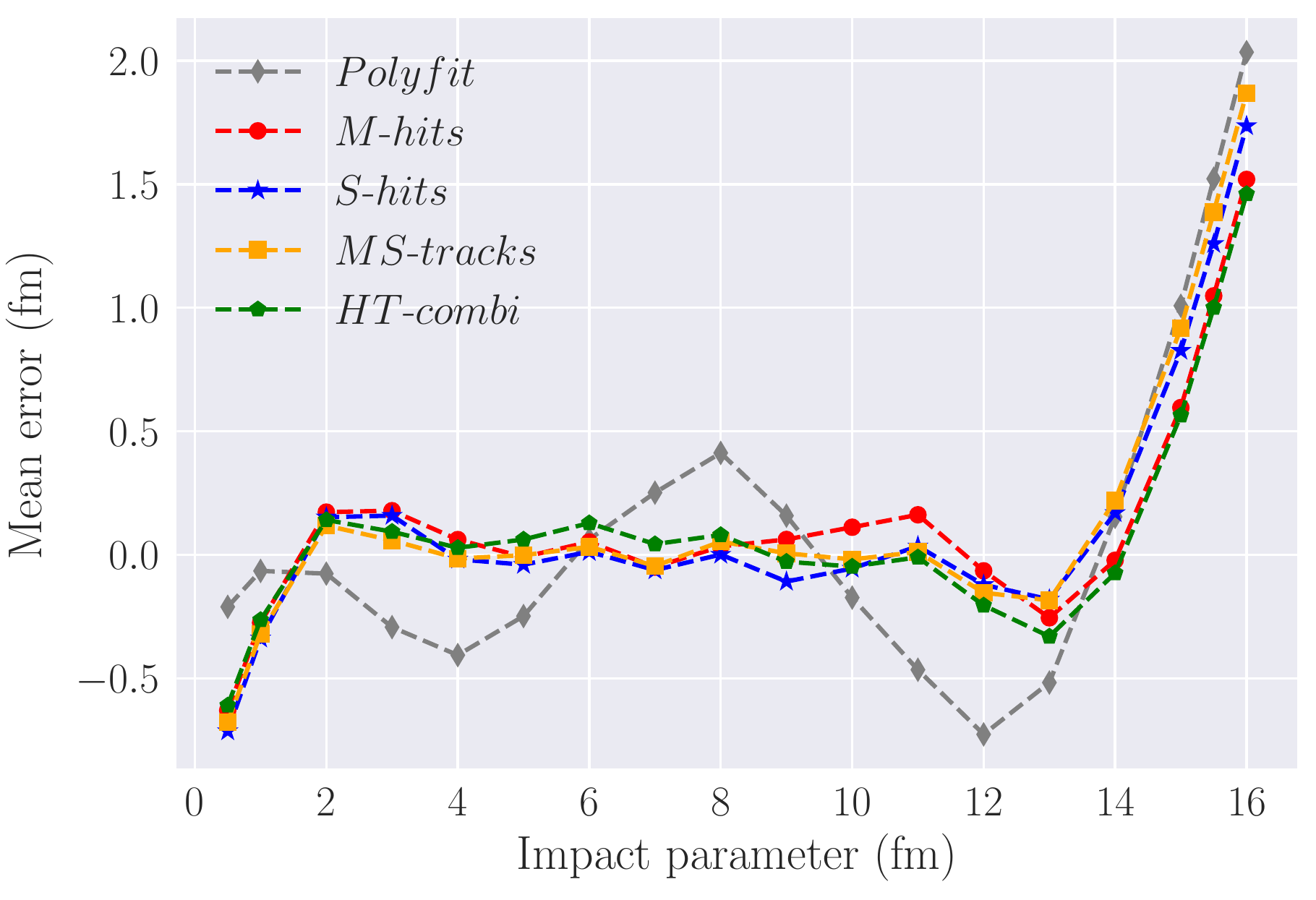}
    \caption{(Color online) Mean error of the predictions as a function of the impact parameter. The events used are from dataset $\textit{Test1}$, and predictions are for a fixed impact parameter. The error bars are smaller than the symbol size.}
    \label{4}
\end{figure}

However, the standard deviation of error in predictions quantifies only the precision of the model. The predictions can be considered both accurate and precise only if the error distributions have a mean close to zero and an acceptable precision. Figure \ref{4} shows the mean of the error in predictions as a function of the impact parameter for \textit{Test1}. The polynomial fit model has a poor accuracy in comparison to DL models, despite its comparable precision in mid-central and peripheral events. The DL models have a mean error between -0.33 to 0.22~fm for events with impact parameter 2-14~fm, while the mean for the \textit{Polyfit} model fluctuates between -0.7 and 0.4~fm. For events in the range 5-14~fm, the $\textit{HT-combi}$  and \textit{Polyfit} offer a relative precision of 4-9~$\%$ and 2-8~$\%$ respectively. Despite their similar precision (for 5-14~fm), $\textit{HT-combi}$  yields more accurate predictions, with a mean error of -0.33 to 0.13~fm, while the polynomial fit exhibits mean errors varying between -0.7 to 0.4~fm. These results indicate that the DL models use more information than just the number of charged tracks to determine the impact parameter.

\begin{figure}
   \centering
   \includegraphics[width=0.485\textwidth]{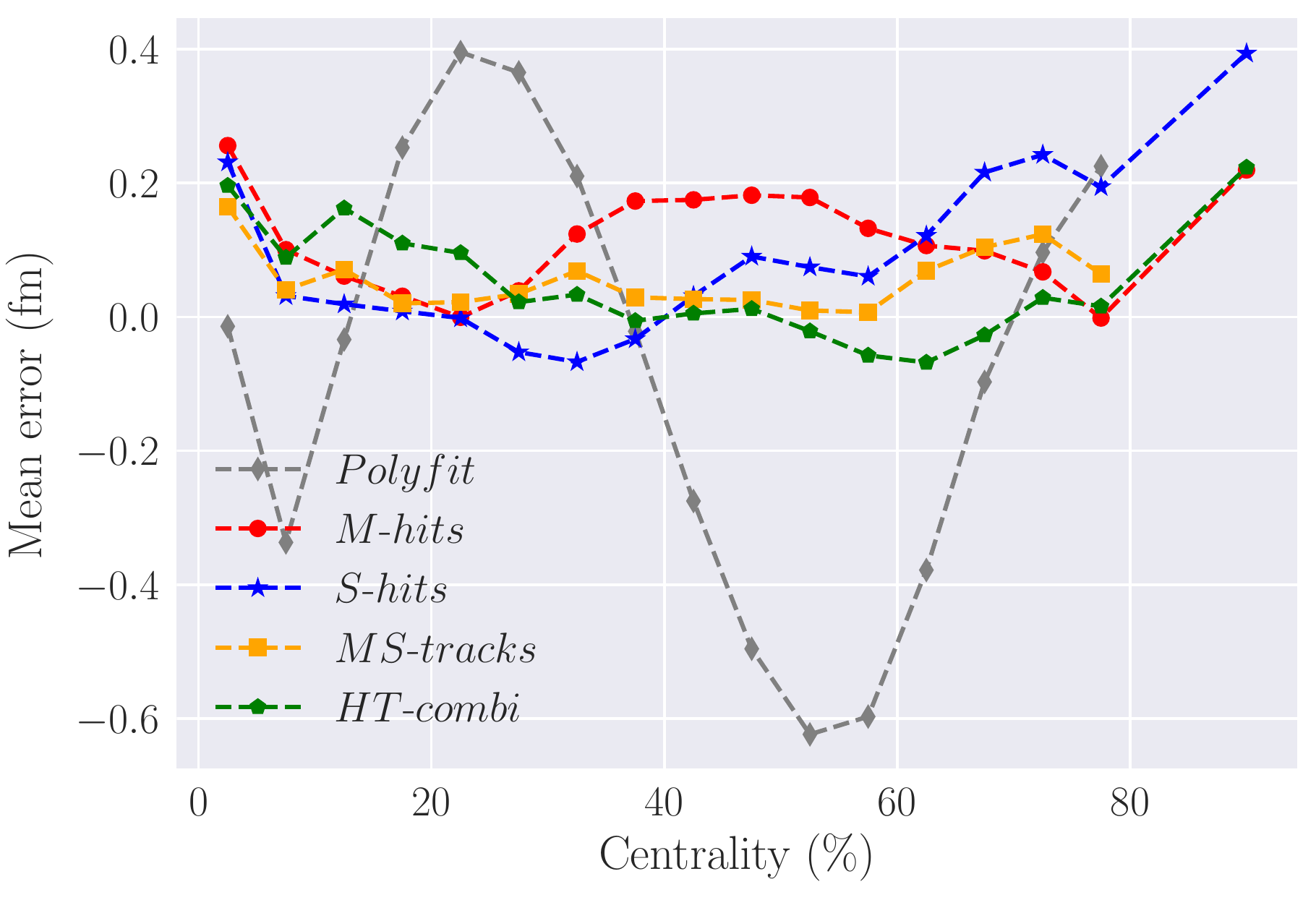}
     \caption{(Color online) Mean error in predictions as a function of centrality. Dataset \textit{Test2} is used in which peripheral events are more likely to occur. The track multiplicity is used for the centrality binning. The points at 90  $\%$ centrality are results from events with no tracks reconstructed. Therefore the \textit{Polyfit} and \textit{MS-Tracks} model do not have a data point at 90 $\%$ centrality.}
     \label{5}
 \end{figure}

In an actual collision experiment, the probability of having events with impact parameter ($b$) is proportional to the impact parameter, which gives a different distribution of the impact parameters than the ones used in the \textit{Train} dataset: i.e. peripheral events are more likely. To study the performance of the DL models in such a scenario, dataset \textit{Test2} was used to predict the impact parameter for different centrality classes with a bin width of $5\%$.
The mean of the prediction error is plotted as a function of centrality in figure \ref{5}.
The DL models have a mean error close to zero for most of the centrality classes while there are large fluctuations in the simple polynomial model. Another interesting factor is that the number of events which has at least 1 hit in the MVD detector but no tracks (using MVD and STS hits) reconstructed were about 10$\%$ of \textit{Test2}. These are ``empty" events for the  
track multiplicity based method. However, the DL models can use hits to make predictions of the impact parameter of these events, though the error is large in comparison to their predictions for central and mid-central events.

\begin{figure}
    \centering
    \includegraphics[width=0.485\textwidth]{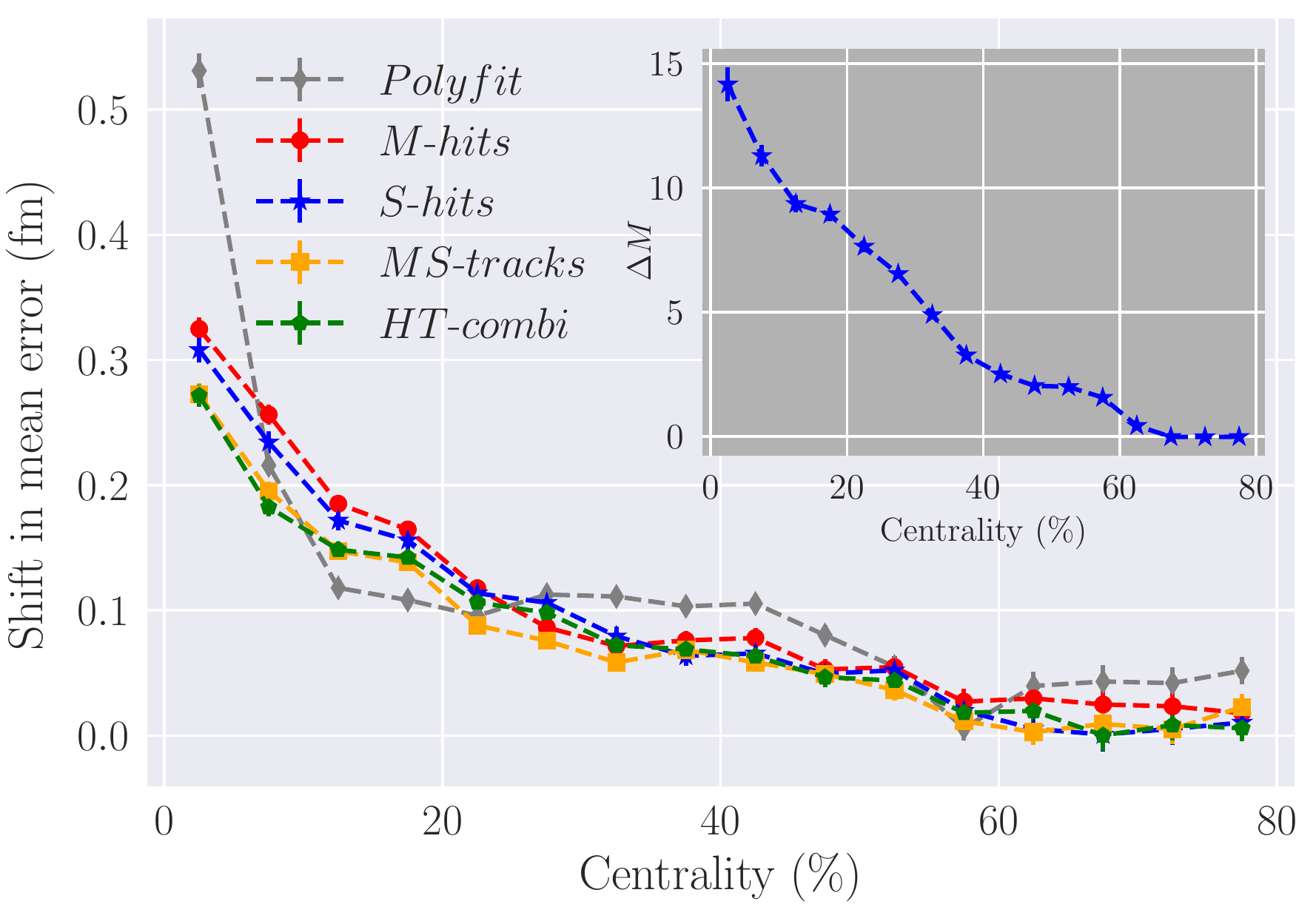}
    \caption{(color online) Inset: Difference of the mean track multiplicity for datasets \textit{Test3} and \textit{Test2} ($\Delta M$) as a function of centrality. The change of the pion cross section in \textit{Test3} is expected to be more visible in central collisions and leads to a larger number of charged tracks. Large figure: Difference of the mean of error distributions for datasets \textit{Test3} and \textit{Test2} as a function of centrality. The increased pion production in central events leads to a systematic under-prediction of the impact parameter in \textit{Test3}. However, the DL models appear less model dependent than the polynomial fit.
    \label{7}}
\end{figure}

The accuracy of the reconstructed impact parameter of an event can depend on how accurate the simulation model can describe the outcome of single events. This introduces a bias on the predictions from the choice of the event generating model. The dependence of the DL predictions on the physics model is studied by predicting the events from a separate dataset, that introduces different physics (\textit{Test3}), on the DL model trained on dataset \textit{Train}. To generate \textit{Test3\textit}, the final charged particle multiplicity in the tested events was modified by an increase of the pion production cross section in UrQMD. To do so, the $\Delta$-baryon absorption cross section in the UrQMD model was decreased by a factor 2, resulting in an increased pion production, especially for central collisions. The increased number of pions is  reflected in the difference of the mean charged track multiplicity ($\Delta M)$ for events in \textit{Test3} and \textit{Test2}, for a given centrality as shown in the inlet of figure \ref{7}. There is a difference of about 14 tracks for most central events and it reduces to less than 3 for peripheral collisions. This change in physics is translated to a shift in the mean of the error distributions ($\mu_{err}^{shift}$) given by,
\begin{equation}
\mu_{err}^{shift}= \sqrt{(\mu_{errT3}- \mu_{errT2})^{2}} 
\end{equation}
where $\mu_{errT3}$ and $\mu_{errT2}$ are the mean in the prediction errors for dataset \textit{Test3} and \textit{Test2} respectively. This shift in mean is plotted as a function of centrality in figure \ref{7}. It is observed that the DL models show a shift in the mean of up to 0.32~fm while the polynomial fit shows a shift up to 0.53~fm. The shift is more evident for central collisions as expected. 
This means that the DL network learns more information than the \textit{Polyfit} about the event features independent of the event multiplicity and thus is less model dependent than a simple fit.
The \textit{MS-tracks} and \textit{HT-combi} show slightly better robustness to the physics modification compared to\textit{ M-hits} and \textit{S-hits} models. The track multiplicity of the event is definitely an important feature with strong correlation with impact parameter. However, as DL models learn other information in the data in addition to track multiplicity, they tend to be more robust than the polynomial fit model which essentially depends only on the track multiplicity.

\section{Conclusion and Discussion}\label{conc}
In this study, we have shown that pointnet-based DL models can be used for an accurate determination of impact parameter in the CBM experiment. The use of input data with minimum preprocessing and high processing speed of DL models make it an ideal candidate for online event selection. It is also interesting to note that all four types of models (\textit{M-hits}, \textit{S-hits}, \textit{MS-tracks} and \textit{HT-combi}) lead essentially to comparable precision in the determination of the underlying impact parameter. Indeed track-based modelling shows only marginally better performance in evaluating validation data.

The DL models are a reliable tool for impact parameter determination over impact parameters in the range 2-14~fm. Events having an impact parameter less than 2~fm is only a very small fraction of the total events in an experiment. Nevertheless, the predictions are still better than the prediction from the polynomial fit which fails for most central events. The deep learning models show a superior performance in comparison to a simple model which relies only on the track multiplicity.  However, all methods to estimate impact parameter will have a bias in the predictions acquired from the physics models used in data generation. This is true for Glauber based estimation as well. In addition, the training data used in this DL study using the UrQMD model and CBM detector simulation may not perfectly represent real data. This model bias can be estimated for DL models by comparing the predictions of a model  on different event generator data. This bias could also be minimised by using events from multiple event generators in the training samples. The use of these DL models in the experiment would also require more investigations into the robustness of the model against expected detector noise and efficiency. However, these are beyond the scope of this paper and are desirable for further investigations in future. The practical application of a DL based event selection algorithm however requires further studies on the scalability of the prediction speed on multiple GPUs and also the possibilities to incorporate other selection criteria. It was also found that the model complexity can be further reduced without significant change in the performance. Therefore, the prediction speed can also be scaled up by reducing the number of model parameters. The results of an ablation study on the \textit {M-hits} model to see the performance change with reduced number of parameters are described in \ref{appendixb}.  

The PointNet based models presented in this study use information like tracks and hits of particle which are available in every heavy ion collision experiment immediately during data collection. It is intended to use the developed model architectures in other heavy ion collision experiments, e.g.: ALICE at the Large Hadron Collider (LHC) or HADES at the SIS18. Here the model can be employed and studied with real data. Moreover, the models used in this paper can readily be generalised for tasks other than impact parameter determination. In the future it is worthwhile to study if a similar model can be used also for more complex tasks like identification of rare physics processes, determination of other observables and the detection of  QCD phase transition.

\section*{Acknowledgement}
The authors thank Volker Friese, Manuel Lorenz, Ilya Selyuzhenkov and Steffen Bass for helpful discussions and comments, Dariusz Miskowiec, Dmytro Kresan and  Florian Uhlig for help with unigen code. We thank the CBM collaboration for the access to the CBM-ROOT simulation software package.
MOK, JS, and KZ thank the Samson AG and the BMBF through the ErUM-Data project for funding. JS and HS thank the Walter Greiner Gesellschaft zur F\"{o}rderung der physikalischen Grundlagenforschung e.V. for her support. MOK thanks HGS-HiRe and GSI through a F$\&$E grant for their support.
Computational resources were provided by the NVIDIA Corporation with the donation of two NVIDIA TITAN Xp GPUs and the Frankfurt Center for Scientific Computing (Goethe-HLR).

\newpage

\appendix
\section{The PointNet Structure}
\label{appendix}
The PointNet model for classification or regression uses two joint alignment networks to transform the data in input and feature space and a symmetric function to accumulate all global features. The joint alignment network makes the model invariant to certain geometric transformations while the symmetric function makes the model invariant to input order.

All the models in this study used ReLU activation units and Adam optimiser (learning rate = 0.00001). The convolution layers were always followed by Batch normalisation layers. The convolution operations (1-D) used kernels of size 1 to ensure that the local features from individual points were segregated separately. Dropout layers with dropout probability of 0.5 were used after every dense layer in the models to control overfitting.
The models use a common structure for input and feature transformation networks ($I_{trans}$ and $F_{trans}$) as illustrated in figure \ref{alignnet}. When used for input transformation, the network has input dimensions N$\times $F where N is the maximum number of hits or tracks (depending on the model) in the data and F is the number of input attributes per point. For the feature transformation network, the input dimensions are N$\times $K where K is the number of  features maps produced by the previous convolution layer. The input passes through a series of convolution and Batch normalisation layers to perform input order independent feature extraction before aggregation of global features using an average pooling layer. The global features are then regressed using a Deep Neural Network (DNN) to output F${^2}$ or K${^2}$ numbers respectively which act as the  transformation matrices. The overall structure of models is also similar to the alignment networks.  Features extracted after input and feature transformations also use average pooling layer to collect global features and then finally find the impact parameter from these features using a DNN. The network architecture and the hyperparameters of the models used in this study are described in detail below.\\

\subsection{Model-1 (M-hits)}
\begin{figure*}[h]
    \centering
    \includegraphics[width=0.9\textwidth]{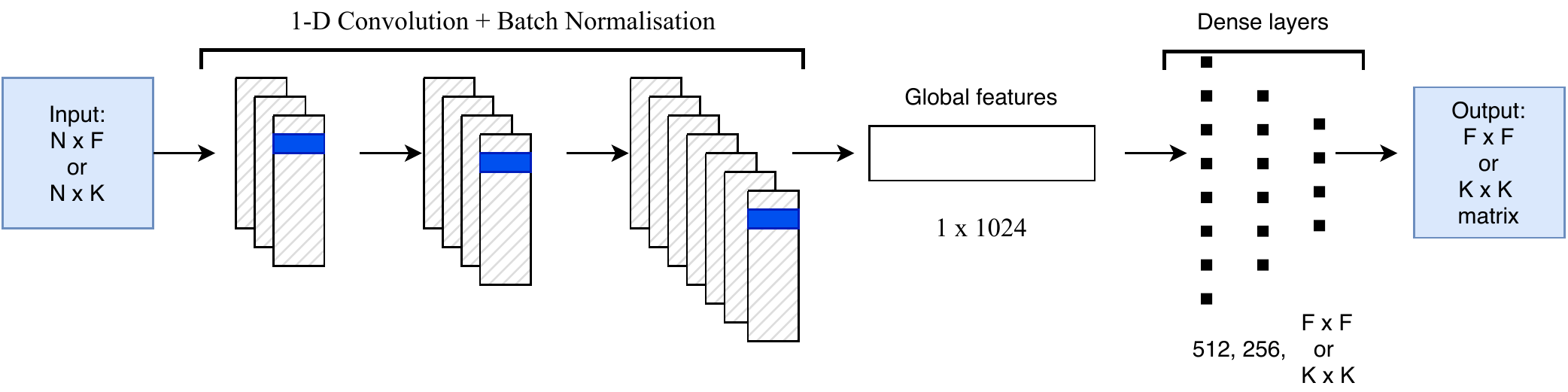}
    \caption{General structure of the joint alignment networks. This network is used as input and feature transformation networks in all the models. All convolution layers are followed by Batch normalisation layers. The convolution kernels (blue rectangles) have size 1. When used as input transformation network, the input has dimensions N$\times$F where N is the maximum number of hits or tracks expected in an event and F is the number of input attributes of each hit or track (eg. F= 3 if the model uses x, y, z coordinates of all hits as input feature). The alignment network then learns an F$\times$F matrix. Similarly, when the network is used as feature transformation network, the input has dimensions N$\times$K where K is the number of features maps produced by the convolution layer preceding the feature transformation network. }
    \label{alignnet}
\end{figure*}

\begin{figure*}[h]
    \centering
    \includegraphics[width=0.9\textwidth]{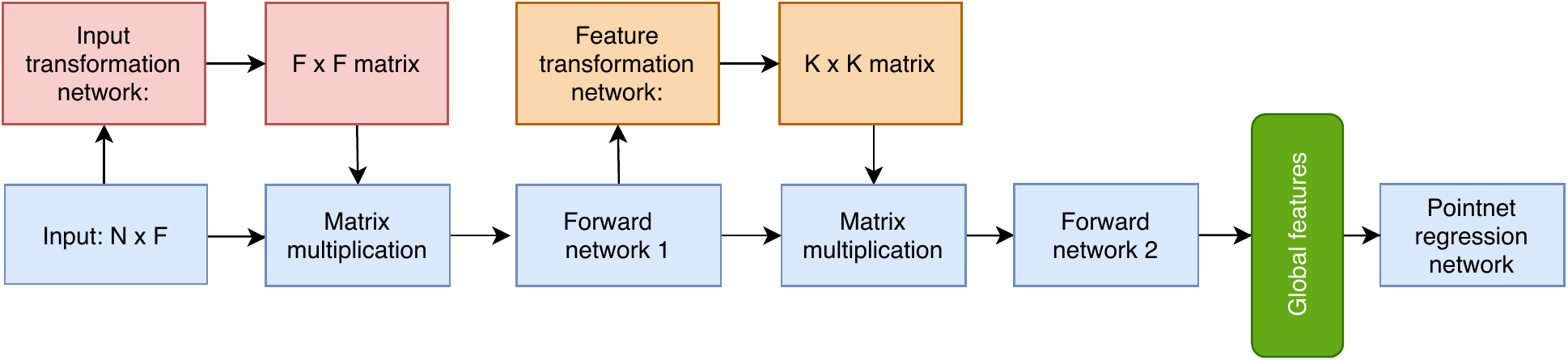}
    \caption{General structure of \textit{M-hits}, \textit{S-hits} and \textit{MS-tracks} models. The main difference among the models is the input shape which depends on the maximum number of hits or tracks expected in an event (N) and the number of attributes considered for each hit or track (F). This changes the dimensions of the input transformation matrix accordingly. The dimensions of the feature transformation matrix (K$\times$K) are equal to the number of feature maps extracted by the last convolution layer of forward network 1.  }
    \label{mhits}
\end{figure*}

The model architecture of \textit{M-hits} which use hits from MVD is illustrated in figure \ref{mhits}. The input transformation network (FIG. \ref{alignnet}) learns to generate a 3$\times$3 matrix (F$\times$F) which transforms the three dimensional points in the input space. This network uses three convolution layers which generates 64, 128 and 1024 feature maps respectively. The transformed input passes through Forward network 1 which consists of 2 layers of convolutions producing 64 feature maps each. This data is then transformed by a 64$\times$64 matrix (K$\times$K) learned by the feature transformation network (FIG. \ref{alignnet}) with three convolution layers (64, 128 and 1024 feature maps). The data then passes through a series of 3 convolution layers (Forward network 2) with 128, 256 and 512 feature maps respectively. Finally the global features are collected using Average pooling function with pool size of 1995. This segregates 512 global features of the event which is passed to the pointcloud regression network made of a three layered Neural Network with 256, 128 and 1 neuron respectively.\\

\subsection{Model-2 (S-hits)}
This model uses hits of particles detected in the STS detector. The structure is similar to \textit{M-hits} model with the only difference happening in the input shape. The input has dimensions 9820$\times$3 as the maximum number of hits present in the training dataset was 9820.\\

\subsection{Model-3 (MS-tracks)}
The basic structure of data flow in this model is similar to that of \textit{M-hits} model. However, this model requires a tracking algorithm as it utilises the tracks of particles reconstructed from STS and MVD as features.
An input transformation network with similar structure to the one in \textit{M-hits} model learns the 12$\times$12 alignment matrix. The forward network 1 comprises of 2 convolution layers producing 128 feature maps each. The extracted features from Forward network 1 are transformed by a 128$\times$128 matrix (K$\times$K). The matrix is learned by the alignment network with 3 convolution layers (128, 256 and 1024 feature maps). The extracted features then pass through a Forward network 2 similar to \textit{M-hits} model. Then an average pooling layer (pool size= 560) segregates the global features and feeds them to a  regression network similar to that of \textit{M-hits}.\\

\subsection{Model-4 (HT-combi)}

\begin{figure*}[h]
    \centering
    \includegraphics[width=0.8\textwidth]{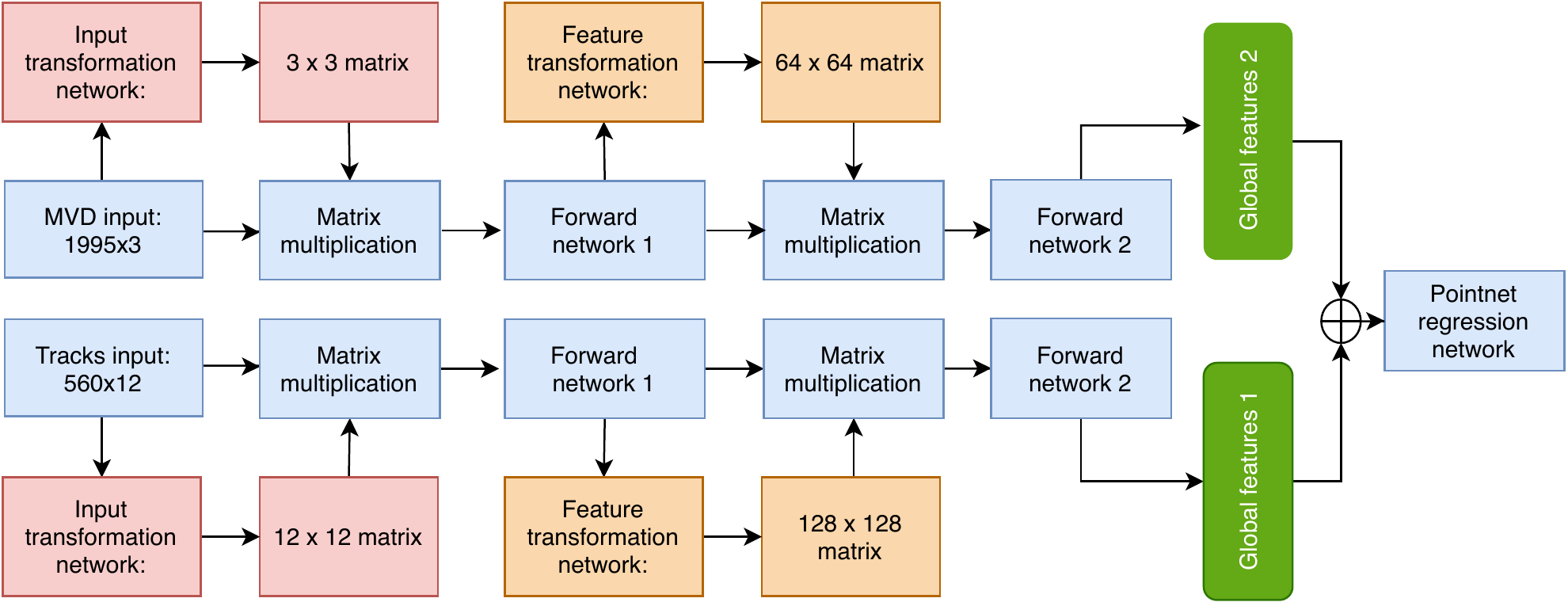}
    \caption{Structure of the model \textit{HT-combi}. The model is a combination of \textit{M-hits} and \textit{MS-tracks}. Both models independently extract the global features and then they are concatenated before being fed to a regression network.}
    \label{thcombi}
\end{figure*}

In this model, two separate networks similar to \textit{M-hits} and \textit{MS-tracks} run parallel to perform input transformation, feature transformation and global feature extraction. Finally, the global features are concatenated and fed into the regression network with 512, 256, 128 and 1 neuron respectively. The model structure is illustrated in figure \ref{thcombi}.\\

\section{Performance without joint alignment networks}
\label{appendixb}
The joint alignment networks used in the study are a straightforward application of those networks in \cite{qi2017pointnet}. The intuition behind the usage of these networks is to enable the DL models to learn certain transformations in the data that can better expose the correlations in the data for the convolution layers after it. The same idea of input transformations can be extended to feature space too. The learned features could be transformed before feeding it to the next convolution layer. If there exist some correlations that are highlighted upon the transformation, the model could benefit from the transformation. We tested the performance change of \textit{M-hits} model without the alignment networks. The results are tabulated in \ref{tablemhit}. It is seen that the performance is only marginally improved when both alignment networks are used. When only the feature transformation network was removed, the validation loss was slightly better than the case with both alignment networks removed. Although the performance improvement is marginal, it must be noted that the models without the alignment networks tend to overfit more compared to the original M-Hits model. This was observed in the difference between their training and validation loss. The use of an alignment network also gives more stability in the training, avoiding sudden fluctuations in the validation loss. Moreover, the model also converges faster with the use of alignment networks.

\begin{table}

\begin{tabular}{p{3.7cm}p{0.99cm}p{0.99cm}p{0.99cm}p{0.7cm}}
\hline
\hline
Model                         & MSE    & MAE    & $R^{2}$     & Epoch \\
\hline
original \textit{M-hits}               & 0.4290 & 0.5123 & 0.9789 & 128   \\
without $I_{trans}$ and $F_{trans}$ & 0.4378 & 0.5196 & 0.9784 & 455   \\
without $F_{trans}$              & 0.4304 & 0.5137 & 0.9788 & 172\\
\hline
\hline
\end{tabular}
\caption{\label{tablemhit} Performance change of the \textit{M-hits} model without the alignment networks.  $I_{trans}$ and $F_{trans}$ are Input transformation and Feature transformation networks respectively. The MSE, MAE and $R^{2}$ are for the validation data . The last column shows the number of epochs the model took to converge to its best performance. }
\end{table}

\end{document}